\documentclass[aps,prl,showpacs,twocolumn,reprint,superscriptaddress]{revtex4}
\usepackage{amsmath}
\usepackage{amssymb}
\usepackage{graphicx}
\usepackage{hyperref}

\input{tcilatex}
\begin{document}

\title{In-plane Charge Fluctuations in Bismuth Sulfide Superconductors}
\author{Anushika Athauda,$^{1}$ Junjie Yang$^{1}$, Seunghun Lee$^{1}$,
Yoshikazu Mizuguchi$^{2}$, Keita Deguchi$^{3}$, Yoshihiko Takano$^{3}$,
Osuke Miura$^{2}$, Despina Louca$^{\ast }$}
\affiliation{Department of Physics, University of Virginia, Charlottesville, VA 22904,
USA. \ \\
$^{2}$Tokyo Metropolitan University, Tokyo,192-0397, Japan.\\
$^{3}$National Institute for Materials Science, Tsukuba 305-0047, Japan.}
\date{\today }

\begin{abstract}
Evidence for local charge fluctuations linked to a charge disproportionation
of the Bi ions in the distorted lattice of superconducting LaO$_{1-x}$F$_{x}$%
BiS$_{2}$ is presented. \ In-plane short-range distortions of sulfur atoms
up to 0.3 \AA\ in magnitude\ break site symmetry and create two distinct
environments around Bi. \ Out-of-plane motion of apical sulfur brings it
closer to the La-O/F doping layer with increasing $x$ that may lead to a
charge transfer conduit between the doping layers and the superconducting BiS%
$_{2}$ planes. \ The mechanism for superconductivity may arise from the
interplay between charge density fluctuations and an enhanced spin-orbit
coupling suggested theoretically, that induces spin polarization.
\end{abstract}

\maketitle

The rich physics of the Bismuth Sulfide (BiS$_{2}$) based superconductors 
\cite{mizuguchi1,mizuguchi2,demura,xing,kuroki,kotegawa,deguchi} includes a
possible charge density wave\ (CDW) instability\cite{yildirim,wan}, atomic
modes leading to phonon softening\cite{yildirim} and spin-orbit (SO)
interactions with a hidden spin polarization \cite{zhang}. \ Much of the
recent interest in superconductors has been on materials whose parent ground
state is magnetic i.e. the Fe and Cu based systems. \ The BiS$_{2}$
superconductors are not magnetic, possibly of the conventional type but with
strong electronic correlations \cite{wang}. \ Density functional theory
calculations suggested that spin polarization is present even though the
crystal lattice is centrosymmetric \cite{zhang,wan}, arising from coupling
of the electron orbital to its spin. \ Such SO coupling has been at the
heart of spintronic physics, exemplified at the interfaces of films like
LaAlO$_{3}$/SrTiO$_{3}$ \cite{hwang}, where a large Rashba SO coupling is
important in stabilizing the superconducting state \cite{caviglia}. \ Given
its quasi-two dimensional crystal nature, BiS$_{2}$ may exhibit SO coupling
through the Dresselhaus effect \cite{zhang}, the 3-dimensional analogue of
the Rashba effect. \ This can happen without breaking crystal inversion
symmetry but rather be driven by atomic site asymmetry \cite{zhang}. \ 

\ In the two families of BiS$_{2}$ superconductors, Bi$_{4}$O$_{4}$(SO$_{4}$)%
$_{1-x}$ \cite{mizuguchi1} and LnO$_{1-x}$F$_{x}$BiS$_{2}$ (Ln = La, Nd, Pr,
Ce, and Yb) \cite{mizuguchi2,yazici,li} discovered thus far, vacancies and
crystal defects are common. \ While the SO$_{4}$ layer carries vacancies in
the former compound, the LnO$_{1-x}$F$_{x}$BiS$_{2}$ system shows a peculiar
relation of its crystal structure to superconductivity. \ The parent
compound shows high crystallinity but upon doping, T$_{C}$ is maximized when
crystal disorder is induced by high pressure annealing. \ The relation of
disorder and superconductivity has been intensely studied in cuprates \cite%
{dsfisher} and, earlier, in amorphous solids \cite{johnson,kapitulnik}. \
Disorder shortens the coherence length, enhances thermal fluctuations and
induces vortex pinning but without long-range ordering into a vortex lattice 
\cite{dsfisher,larkin}. \ Such a state has been argued to correspond to a
vortex glass \cite{mpafisher}. \ However, the structural disorder observed
in LnO$_{1-x}$F$_{x}$BiS$_{2}$ is different. 

In this system, the dominant carriers are from the Bi \textit{6p} orbitals
that strongly hybridize with the S \textit{3p} near the Fermi surface. \ A
CDW instability arises from a ($\pi $,$\pi $,0) mode which corresponds to
in-plane displacements of S atoms (0.18 \AA ) around the M point\cite{wan}.
\ At the same time, SO coupling has a considerable effect on the electronic
band structure around the X point of the Brillouin zone. \ Without SO
coupling, four bands cross below E$_{F}$ in the optimally doped composition,
x $\sim 0.5$. \ With SO coupling, only two bands cross below E$_{F}$. \ SO
coupling originates from the inversion asymmetry and creates an effective
magnetic field that lifts the spin degeneracy\cite{liu}. \ Are the new
Bi-based superconductors conventional BCS type or a new class of non-BCS
type, yielding a new route to unconventional superconductivity? \ Here, we
provide evidence for the latter by analyzing neutron scattering data. \ Our
results provide the experimental evidence for in-plane short-range S
distortions ($\sim 0.3$ \AA\ in magnitude) that lead to charge fluctuations
around Bi. \ Crystal strain develops along the compressed \textit{c}-axis
but the out-of-plane S atoms move closer to the doping layers. \ The local
distortions present in this system may be an experimental realization of a
vortex pinning mechanism in the glassy state of a superconductor. \ 

The quasi two-dimensional layered crystal structure of LaOBiS$_{2}$ (see ref 
\cite{mizuguchi1,mizuguchi2,demura} for sample preparation) shown in Fig.
1(a) consists of the superconducting BiS$_{2}$ bilayers sandwiched between
insulating La$_{2}$O$_{2}$ blocking layers. The Bi ions form a square
pyramidal edge shared lattice, coordinated by four in-plane (\textit{S1})
and one out-of-plane (\textit{S2}) sulfur ions. The plane is buckled with a
S1-Bi-S1 bond angle of $\sim $172.1$^{o}$. In the tetragonal symmetry, the
base of the pyramid is a perfect square, and the Bi-S1 bonds are of the same
length, 2.876 \AA . \ The out-of-plane Bi-S2 bond length of 2.476 \AA\ is
significantly shorter. The substitution of oxygen (O$^{2-}$) with fluorine (F%
$^{-}$) in the blocking layers introduces electron carriers, and while the
crystal symmetry, \textit{P4/nmm}, does not change, the c-axis contracts by
4 \% and the a-axis expands by less than 0.2 \%. \ The samples become
superconducting under high pressure annealing. \ The BiS$_{2}$ layer
buckling is reversed with doping, in which the Bi ion moves away from the S2
atom, and the Bi-S2 bond length increases to 2.65 \AA\ by \textit{x} = 0.5.
\ The S1-Bi-S1 bond angle becomes 184.8$^{o}$ with a Bi-S1 bond of 2.87 \AA %
. This contradicts theoretical predictions that expect the lattice to become
flat with doping \cite{miura}.

The centrosymmetric space group does not change with doping \cite{mizuguchi2}%
, even though significant Bragg peak broadening along with diffuse
scattering are observed in the neutron diffraction pattern (see Fig. 1(b)
which is a comparison of the x = 0.0 and x = 0.5). \ It is perplexing how
such a distorted crystal lattice is conducive to superconductivity. \
Earlier studies showed that the broadening affects Bragg peaks with nonzero 
\textit{l-}indices while Bragg peaks with \textit{l} = 0 are much sharper as
reported in Ref. \cite{jlee}. \ Fig. 1(d) is a comparison of two nuclear
Bragg peaks, the (110) and (114). The (114) Bragg peak and others with
nonzero \textit{l} become quite broad in the high pressure treated samples.
\ This has been attributed to \textit{c}-axis strain \cite{kajitani},
arising from stacking faults in the presence of partial dislocations from
imperfect stacking of crystal planes.

To investigate the origin of the broadening and diffuse scattering, here we
use a real-space approach to extract the atom specific characteristics of
the distortions and their real-space configuration. Bragg peak broadening
and diffuse scattering can arise from other form of defects besides stacking
faults: the size mismatch between O and F can contribute to Huang
scattering, for instance, while thermal diffuse scattering (TDS) due to
phonons and static displacements of ions can additionally contribute to the
diffuse scattering. \ We previously estimated the Debye temperature to be
around 200 K for the x = 0.5 \cite{anushika} which indicates that TDS may
not be insignificant in the temperatures of interest here, \TEXTsymbol{<} 10
K. \ The real-space atomic structure obtained by Fourier transforming the
total structure factor that includes both Bragg and diffuse scattering is
represented in the pair density function (PDF), $\rho \left( r\right) $. \
The $\rho \left( r\right) $ is plotted in Fig. 2(a) for four compositions. \
This analysis provides the local arrangement of atoms and is very sensitive
to short-range distortions. \ Further details can be found in Refs: \cite%
{louca1,louca2}. \ The $\rho \left( r\right) $ consists of correlation peaks
that correspond to the probability of finding a particular pair of atoms, as
labelled in the figure, at a given distance in space. \ 

While in reciprocal space Bragg peak broadening is very pronounced as seen
in Fig. 1(c), in real space, the PDF correlation peaks change shape with 
\textit{x}, an unusual effect that indicates a local rearrangement is taking
place that may be driven sterically and/or energetically. \ The crystal
symmetry is invariant under doping and the differences observed in the PDFs
of LaO$_{1-x}$F$_{x}$BiS$_{2}$ cannot be explained by the substitution of F$%
^{-}$ for O$^{2-}$ as the two have very similar neutron scattering lengths
and nominal ionic sizes. \ Starting with the parent compound, the PDF data
is compared to a model $\rho (r)_{\func{mod}}$, calculated using the unit
cell dimensions and atomic coordinates of the average structure. \ In Fig.
2(b), $\rho (r)_{\exp }$ (symbols) is compared to the model $\rho (r)_{\func{%
mod}\text{ }}$(dash blue line - average). The overall agreement between the
two curves is good, however very clear differences are observed especially
around 2.5 -3.5 \AA\ that are not due to systematic or statistical errors. \
Some ripples can be seen around the first PDF peak due to termination
effects resulting from the truncation of the Fourier transform at a finite
momentum transfer but the aforementioned differences are well above the
error. \ 

In-plane S1 displacements can reproduce the local configuration that best
fits the $\rho (r)_{\exp }$ of Fig. 2(b). \ The displacement modes are shown
in Figs. 2(c) and 2(d). \ In 2(c), the displacement mode is of the breathing
type. \ Only S1 is displaced while all other atom coordinates are kept the
same as in\ the average structure. \ A new $\rho (r)_{\func{mod}}$ is
calculated using the breathing mode model and is compared to the $\rho
(r)_{\exp }$ also in Fig. 2(b) (solid red line - local). \ In spite of its
simplicity, this model fits the data quite well, especially\ in the 2.5 to
3.5 \AA\ range. \ The breathing mode shown in Fig. 2(c) creates short and
long Bi-S1 bonds that affect the charge distribution around Bi. \ The
distortion creates a double-well displacement potential of the Bi-S1 bonds.
\ The data can be reproduced equally well using the displacement mode of
Fig. 2(d). \ In this model, the S1 atoms are displaced in a ferrodistortive
type mode, and create short and long Bi-S1 bonds in an asymmetric way. \
Both modes are equally likely as they give rise to identical magnitude of
bonds. \ It is conceivable that both the ferrodistortive and breathing type
modes are locally present. \ The asymmetric coordination is reminiscent of
the environment of Bi in Cu$_{4}$Bi$_{5}$S$_{10}$ \cite{olsen} due to the
presence of a lone pair.

In the superconducting state, similar local modes are likely. \ The $\rho
(r)_{\exp }$ of \textit{x} = 0.5 is shown in Fig. 3(a) and is compared to
the $\rho (r)_{\func{mod}}$ for the average model of x = 0.5 (dash blue
line). \ Differences are observed between the two in the same region of
space as in the parent compound even though the shape of the PDF peaks are
different. \ Just as in the parent compound, the breathing and
ferrodistortive modes can fit the data quite well ($\rho (r)_{\func{mod}}$'s
shown in solid lines) but the magnitude of distortion is different. \ Shown
in the inset of Fig. 3(a) are the Bi-S1 and Bi-S2 partial PDFs obtained from
the local and average models for \textit{x} = 0 and \textit{x} = 0.5. \ It
can clearly be seen that the average Bi-S1 bond of 2.8 \AA\ is split to two
locally in both the x = 0 and 0.5 while the split is larger in the x = 0
than in x = 0.5 sample. A third small peak is observed in close proximity to
this bond that is from Bi-S1 bond lengths across planes. \ 

The charge fluctuation inferred from the long and short Bi-S1 bond lengths
is present in all compositions leading to the \textit{x} = 0.50 which is the
only superconducting composition in our series. \ With doping, the Bi - S1
bonds differ in length as shown in the schematic of the Bi tetrahedra of
Fig. 3(b).\ We can also see that upon entering the superconducting phase,
the apical S2 atom gets further away from the planes and closer to the
La(O/F) charge layers. \ The S2 atom can mediate the transfer of charge from
the doping layers to the conduction BiS$_{2}$ layers just like the chain
oxygen does in YBa$_{2}$Cu$_{3}$O$_{7-\delta }$. \ With temperature, the
split of the Bi-S1 bonds in the x = 0.5 changes little except on approaching
T$_{C}$ where the split becomes larger (see Fig. 4(a)). \ On the other hand,
it is clear that with cooling, the height of the S2 atom gets closer to the
donor planes (Fig. 4(b)). \ This is also consistent with the temperature
dependence of S2 in the average structure.

The long and short Bi-S1 in-plane bonds indicate a variance in the Bi
valence due to charge disproportionation. \ The proposed charge fluctuations
are shown in the schematic of Fig. 3(c) for the BiS$_{2}$ planes. \ Two
displacement modes of the breathing type are tested: given that there are
two planes at $z$ $\sim 1/3$ and $z$ $\sim 2/3$, in one mode the S1
distortions are in the (x, -x) direction in each plane respectively, while
in the other, the distortions are in the (x, -y) direction in each plane. \
Both modes yield the same local bond order but might be different
energetically. \ Note that within one unit cell, the two BiS$_{2}$ planes
are rotated by 90 degrees. \ The (x, -x) only breaks the 4-fold rotational
symmetry\ to a 2-fold while the inversion symmetry remains. \ The (x, -y)
distortions break the 4-fold and 2-fold rotational as well as the inversion
symmetries.

The superconducting properties depend on structural defects and on their
arrangement, and knowing the average or ideal structure is not sufficient 
\cite{jorgensen}. \ What can the local structure tell us about the crystal
state of this new superconductor? \ It is clear that superconductivity is
enhanced with the quenched disorder in this system and S1 and S2 distortions
are important. \ Theoretical works have shown that a vortex-glass
superconductor may be present in the bulk of disorder superconductors. In
the presence of disorder, superconducting vortices are pinned, preventing
the formation of vortex lattices \cite{dsfisher}. \ Thus even though the BiS$%
_{2}$ planes are distorted in the non-superconducting and superconducting
samples, long-range order exists in the former as evidenced from the Bragg
structure, and in the absence of pinning centers, the system can be in a
vortex fluid phase with strong fluctuations preventing pairing. \ On the
other hand, in the superconducting sample, the disorder is strong not only
due to the distortions of the BiS$_{2}$ planes but also due to the c-axis
strain and stacking faults, allowing the system to enter the vortex glass
thermodynamic state. \ The stripes of distortions proposed in Fig. 3(c) are 
\textit{short-range} and the presence of stacking faults break the
periodicity thus preventing any long-range order. \ It is possible that such
modes are the cause of vortex pinning. \ 

The doping of F$^{-}$ for O$^{2-}$ changes the oxidation state of Bi
nominally from 3+ to 2+ while introducing electron carriers. \ A simple
estimate of the Bi$^{3+}$ - S$^{2-}$ bond length in the x = 0 yields a value
of $\sim 2.8$ \AA\ that is close to the bond length determined from the
average structure. \ However, \textit{locally}, the Bi$^{3+}$ - S$^{-2}$
bond lengths are split, as shown above. \ We cannot distinguish whether this
is the result of a breathing or a ferrodistortive type mode because they
yield the same\ bond magnitude. \ The leading instability mode consists of
displacements of S1 in the superconducting plane and out-of-plane S2
distortions. \ The in-plane distortions get smaller in the high pressure
annealed x = 0.5 superconducting samples while the out-of plane z-motion of
S2 increases. \ The two proposed modes will both create unequal charge
distributions. \ However, such charge fluctuations are local and cannot
produce a charge density wave. \ Even though this system is not magnetic,
theoretical calculations showed that SO coupling leads to a hidden spin
polarization that has a considerable effect on the electronic band structure
around the X-point of the Brillouin zone. \ How the instability mode
proposed here affects spin polarization will be interesting to investigate.\
\ In general, superconductivity in the presence of strong structural
disorder might be a playing field for discovering exotic states.

\begin{acknowledgments}
The authors would like to thank X. Zhang, M. Hermely and D. Dessau for
useful discussions and B. Li for help with the data reduction process. \
This work has been supported by the National Science Foundation, Grant
number DMR-1404994. \ Work at ORNL was supported by the US Department of
Energy, Office of Basic Energy Sciences, Materials Sciences and Engineering
Division and Scientific User Facilities Division.
\end{acknowledgments}

*To whom correspondence should be addressed

louca@virginia.edu

\textbf{Fig. 1. \ (a) } The crystal structure of LaO$_{1-x}$F$_{x}$BiS$_{2}$
with the BiS tetrahedra. (\textbf{b}) The neutron powder diffraction pattern
(black symbols) of LaOBiS$_{2}$ at 6 K was collected using NOMAD at the
Spallation Neutron Source of Oak Ridge National Laboratory. \ The red, green
and blue solid lines represent the calculated intensity, background and the
difference between the observed and calculated intensities, respectively. (%
\textbf{c}) The neutron powder diffraction pattern of LaO$_{0.5}$F$_{0.5}$BiS%
$_{2}$ at 2 K has its intensity considerably reduced in comparison to the
parent compound of \textbf{(b)}. \ At the same time, the background (green
solid line) has increased. \ (\textbf{d}) The comparison of the (110) and
(114) Bragg peaks between the x = 0 and x = 0.5 compositions shows
significant changes of the non-zero \textit{l} peaks with \textit{x}.

\noindent \textbf{Fig. 2. \ (a)} A plot of the composition dependence of the
data corresponding to the local structures of x = 0, 0.2, 0.3 and 0.5. (%
\textbf{b}) The PDF of the local structure for the parent compound at 6 K
(black symbols) is compared to the average (dashed blue line) and local
(solid red line) model PDFs. The agreement factor (A$_{factor}$) between the
local model and data is 0.156 while between the average model and data it is
0.283. \ \textbf{(c)} A schematic of the BiS$_{2}$ plane showing the
breathing type mode. \ \textbf{(d)} A schematic of the BiS$_{2}$ plane
showing the ferro-distortive type mode. \ The displacements of S1 are either
in the \textit{x}- or \textit{y}-direction.

\noindent \textbf{Fig. 3. \ (a) }The PDF of the local structure for x = 0.5
at 2 K (black symbols) is compared to the average (dashed blue line) and
local (solid red line) model PDFs. \ The A$_{factor}$ between the local
model and data is 0.18 while between the average model and data it is 0.239.
The inset is a plot of the partial PDFs for Bi-S1 and Bi-S2 for x = 0 (top)
and x = 0.5 (bottom). \ The partial of Bi-S1 for both the average and local
models is shown while the partial of Bi-S2 is only shown for the local
model. (\textbf{b}) A schematic of the Bi-S tetrahedron with the bond
lengths determined from the refinement of the local structures of 0 $\leq $
x $\leq $ 0.5. (\textbf{c}) Stripes of charge fluctuations in the two BiS$%
_{2}$ planes of the crystal structure. \ The (x,-x) and (x,-y) refer to\ the
S1 coordinates in the z $\sim 1/3$ and z $\sim 2/3$ planes, respectively.

\textbf{Fig. 4. \ (a) }The temperature dependence of the short (square) and
long (circle) local Bi-S1 bond lengths is contrasted to the Bi-S1 average
bond length (triangle). \ \textbf{(b)} The temperature dependence of the S2
atom from the oxygen plane obtained from fitting the data and from the
Rietveld refinement of the average structure. \ 

\clearpage

\begin{figure}
	\centering
	\includegraphics[width=1\textwidth]{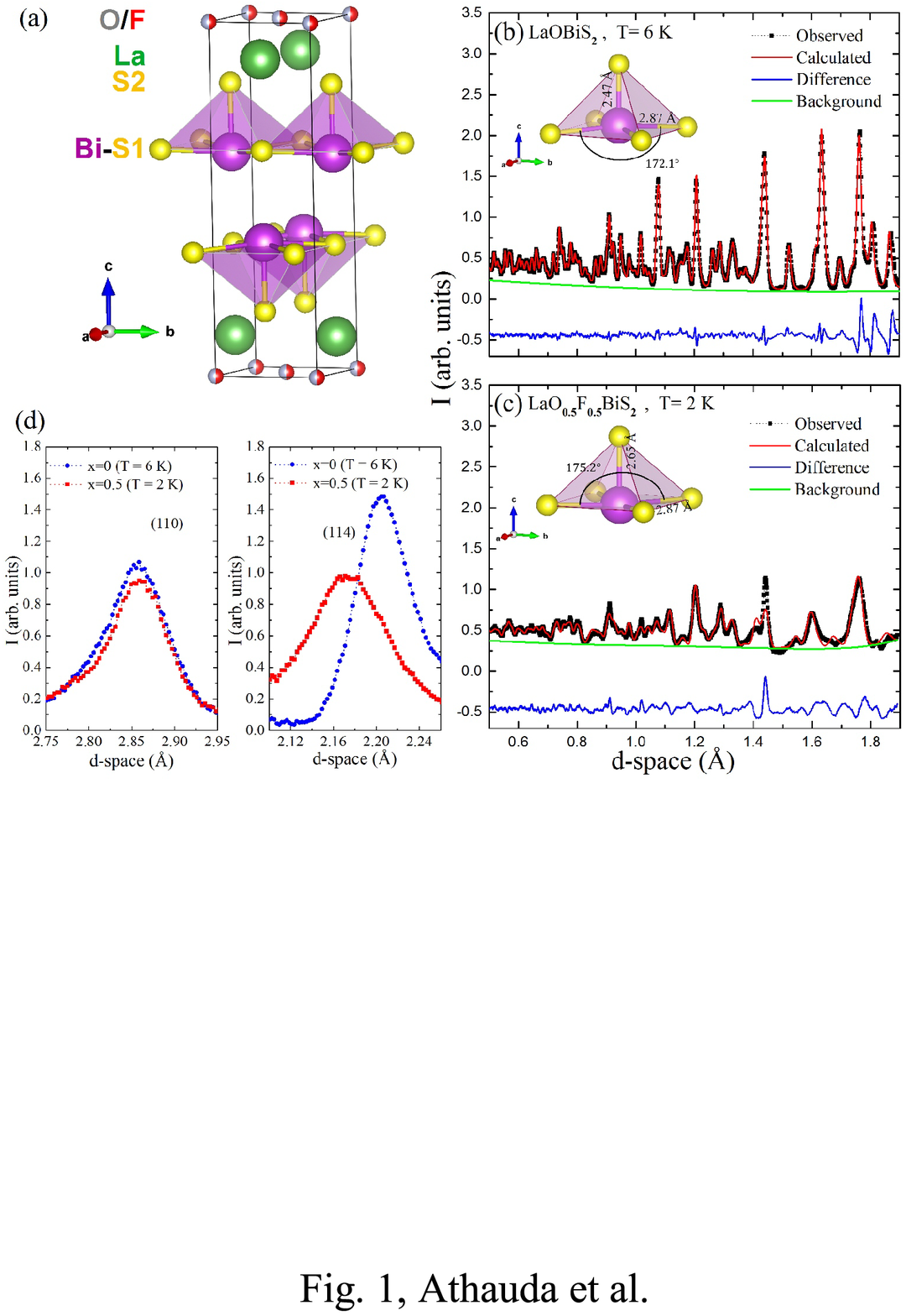}
\end{figure}

\clearpage

\begin{figure}
	\centering
	\includegraphics[width=1\textwidth]{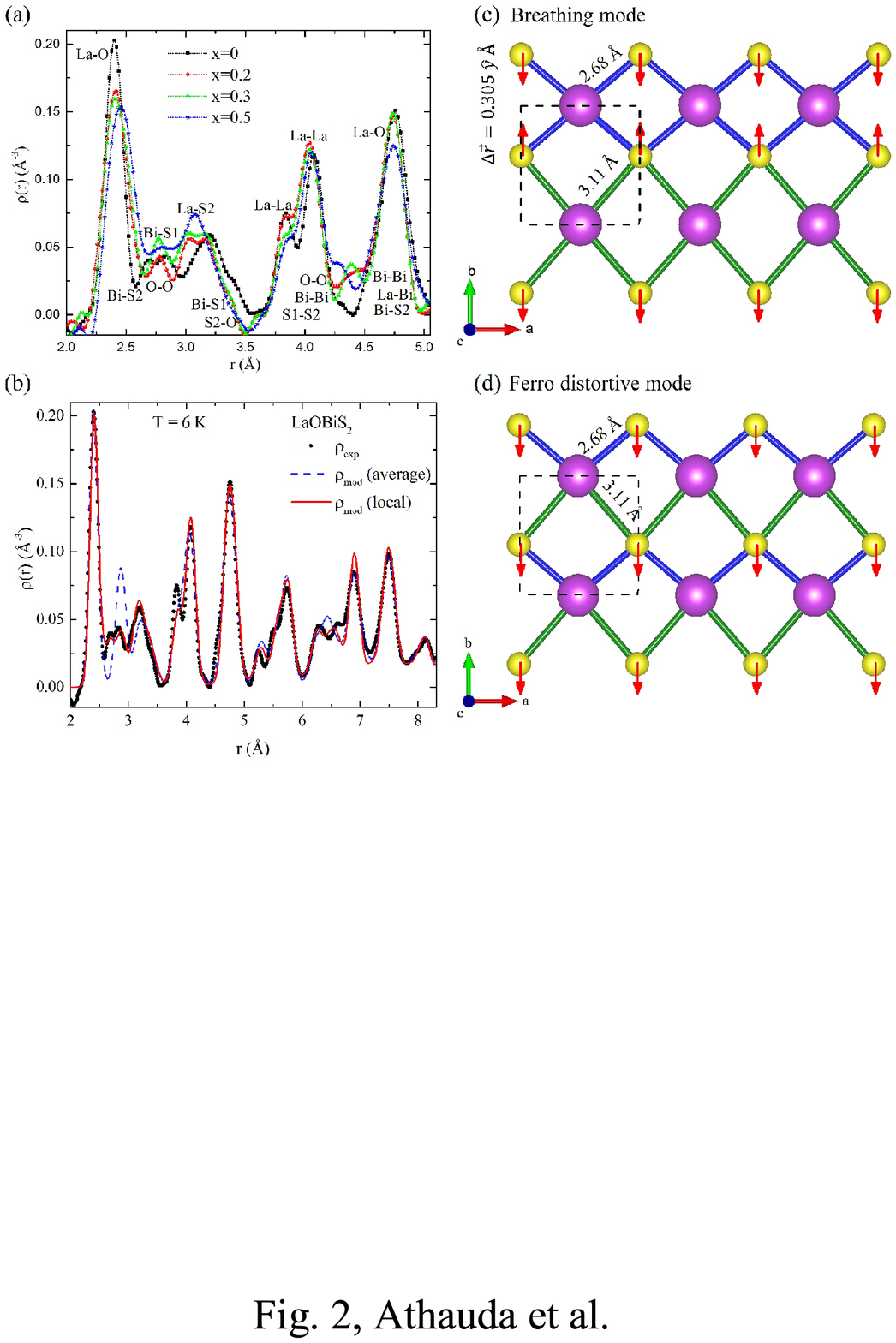}
\end{figure}

\clearpage

\begin{figure}
	\centering
	\includegraphics[width=1\textwidth]{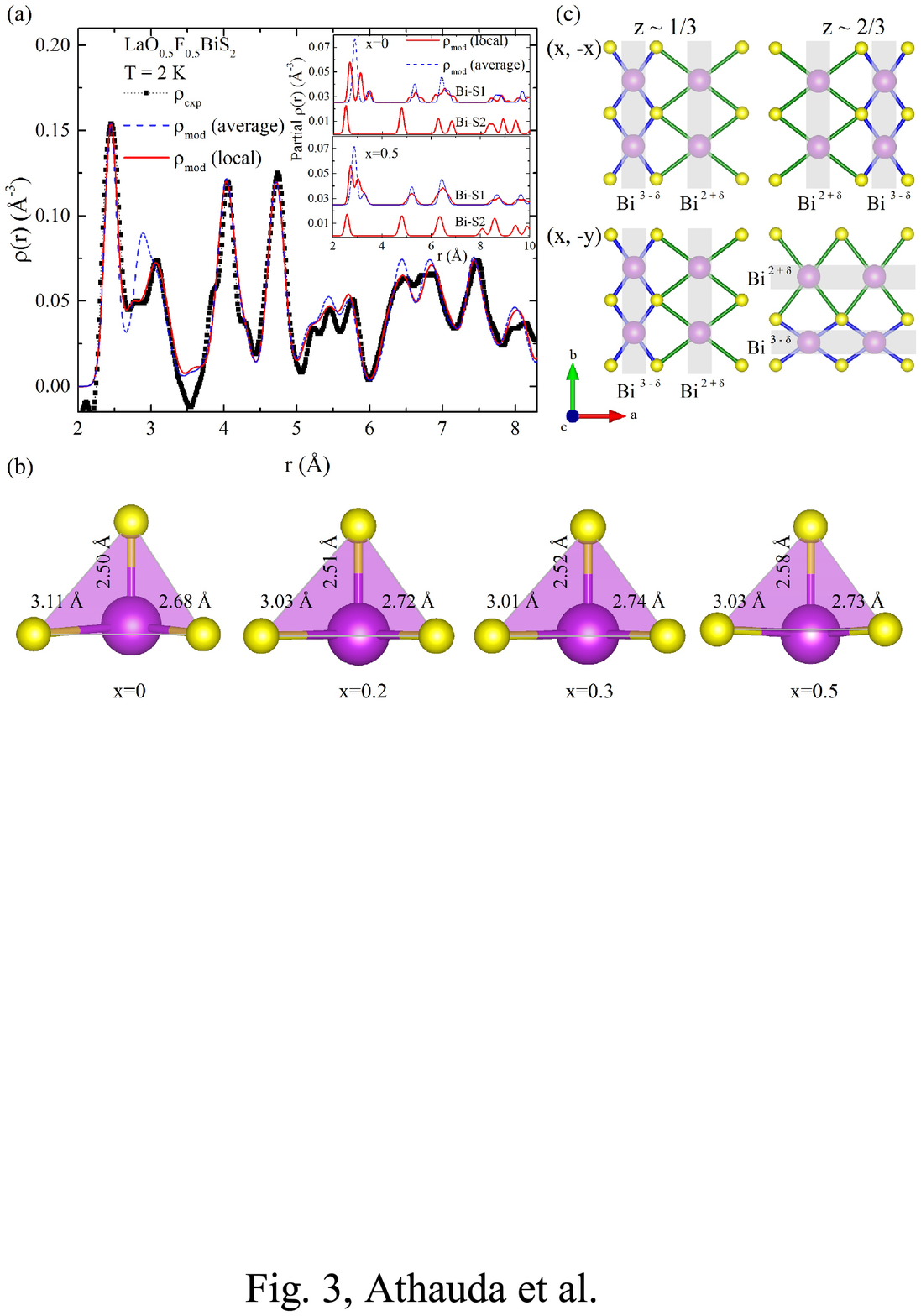}
\end{figure}

\clearpage

\begin{figure}
	\centering
	\includegraphics[width=1\textwidth]{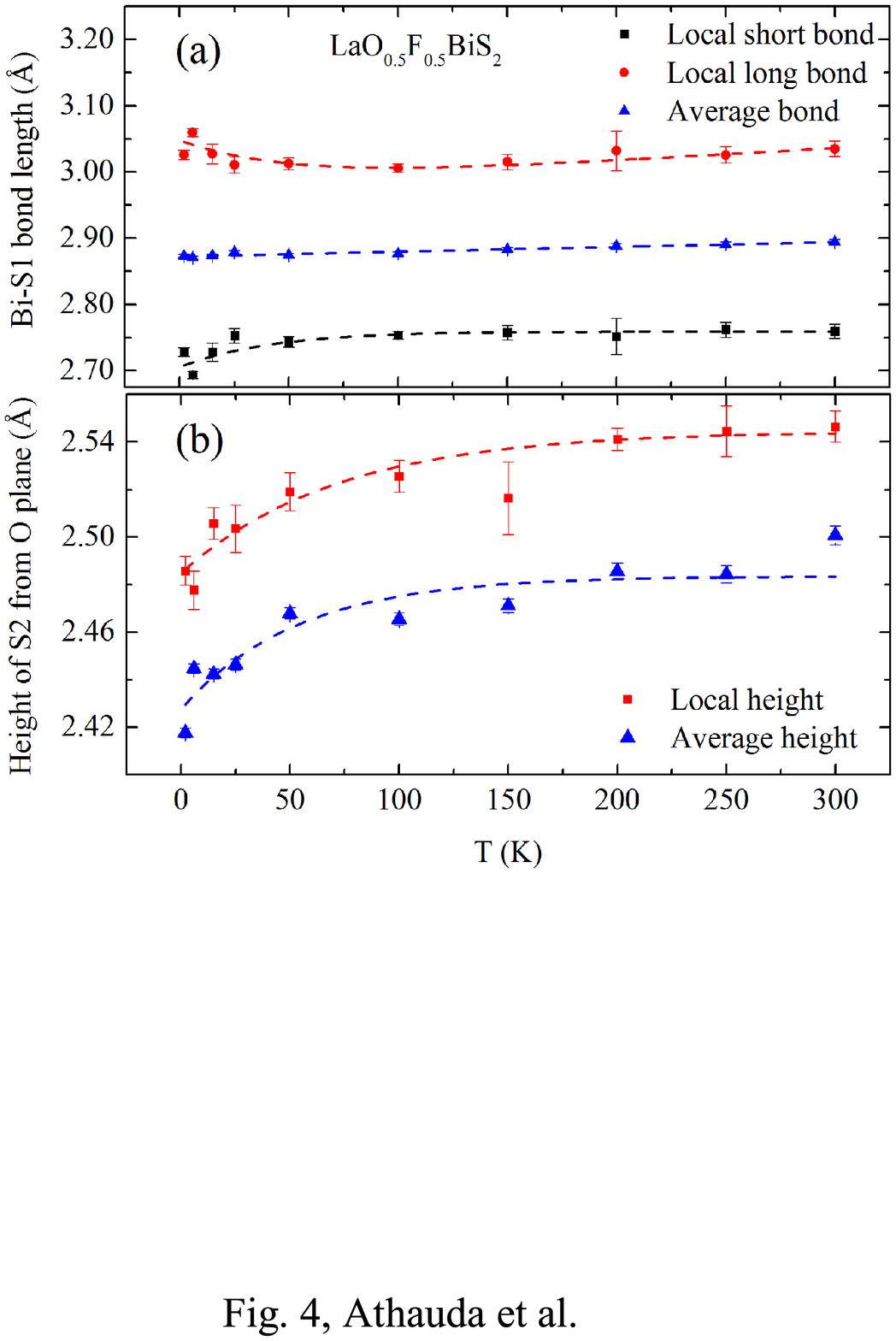}
\end{figure}

\end{document}